\RequirePackage[running]{lineno}

\documentclass[aps,prl,twocolumn,nofootinbib,superscriptaddress,letterpaper,amsmath,amssymb]{revtex4}
\pdfoutput=1

\usepackage{graphicx}  
\usepackage{dcolumn}   
\usepackage{bbm}        
\usepackage{amsmath}
\usepackage{amsfonts}
\usepackage{amssymb}   
\usepackage{epstopdf}
\usepackage{slashed}
\usepackage{hyperref}
\usepackage{multirow}
\usepackage{color}
\usepackage{float}
\usepackage{subcaption}

\usepackage{verbatim}

\usepackage[compat=1.1.0]{tikz-feynman}

\def\gsim{\lower0.5ex\hbox{$\:\buildrel >\over\sim\:$}}
\def\lsim{\lower0.5ex\hbox{$\:\buildrel <\over\sim\:$}}

\newcommand{\be}{\begin{equation}}
\newcommand{\ee}{\end{equation}}
\newcommand{\bea}{\begin{eqnarray}}
\newcommand{\eea}{\end{eqnarray}}

\newcommand{\nbox}{{\,\lower0.9pt\vbox{\hrule \hbox{\vrule height 0.2 cm
\hskip 0.2 cm \vrule height 0.2 cm}\hrule}\,}}

\def\sub#1{_{\lower.25ex\hbox{$\scriptstyle#1$}}}

\newskip\zatskip \zatskip=0pt plus0pt minus0pt
\def\matth{\mathsurround=0pt}
\def\lsim{\mathrel{\mathpalette\atversim<}}
\def\gsim{\mathrel{\mathpalette\atversim>}}
\def\sigv{\ifmmode \langle\sigma v\rangle\else $\langle\sigma v\rangle$\fi}
\newskip\zatskip \zatskip=0pt plus0pt minus0pt
\def\matth{\mathsurround=0pt}
\def\lsim{\mathrel{\mathpalette\atversim<}}
\def\gsim{\mathrel{\mathpalette\atversim>}}
\def\atversim#1#2{\lower0.7ex\vbox{\baselineskip\zatskip\lineskip\zatskip
  \lineskiplimit
  0pt\ialign{$\matth#1\hfil##\hfil$\crcr#2\crcr\sim\crcr}}}

\begin{document}

\thispagestyle{empty}

\vspace{0.5in}

\title{New Physics in Single Resonant Top Quarks}
\author{Shelley Tong}
\author{James Corcoran}
\author{Max Fieg}
\author{Michael Fenton}
\author{Daniel Whiteson}
\affiliation{Department of Physics \& Astronomy, University of California, Irvine}
\begin{abstract}
Searches for new physics in the top quark sector are of great theoretical interest, yet some powerful avenues for discovery remain unexplored. We characterize the expected statistical power of the LHC dataset to constrain the single production of heavy top partners $T$ decaying to a top quark and a photon or a top quark and a gluon. We describe an effective interaction which could generate such production, though the limits apply to a range of theoretical models. We find sensitivity to cross sections in the $10^{2}-10^{5}$ fb range, for $T$ masses between 300 and 1000 GeV, depending on decay mode.
\end{abstract}
\maketitle

\section{Introduction}

A central goal of experiments at the energy frontier is the discovery of new particles.  Many searches for new particles are motivated by theoretical issues with the Standard Model (SM). A complementary approach is to perform searches with significant discovery potential regardless of current theoretical interest~\cite{Craig:2016rqv,Kim:2019rhy}. 

Searches for new physics in interactions involving top quarks benefit from both motivations. As the heaviest (known) fundamental particle, the top quark may play a special role in electro-weak symmetry breaking and have exotic vector-like fermion partners, making an understanding of its production, properties and decay modes of great theoretical interest.  At the same time, the multiple potential decay modes of the heavy top partner $T$ provide many potential discovery modes involving top quarks.

Searches at ATLAS and CMS have mainly studied $T$ decays featuring heavy SM bosons, $T\rightarrow Wb,Zt, Ht$~\cite{ATLAS:2018ziw,ATLAS:2017nap,CMS:2023agg,CMS:2022yxp,CMS:2020ttz,Roy:2020fqf,ATLAS:2022ozf,ATLAS:2023pja,ATLAS:2022tla}  and the many possible decay modes which follow. Significantly less attention has been given to possible modes $T\rightarrow gt,\gamma t$~\cite{Kim:2018mks}, and where this has been studied it has focused primarily on cases in which the exotic heavy fermion is pair produced and one top quark decays leptonically and one hadronically~\cite{Alhazmi:2018whk}, or where a top partner $T$ is produced in association with a top quark, leading to a final state similar to top quark pair production~\cite{Kim:2018mks}.  In order to fully exploit the physics potential of the LHC, it is important to explore all production and decay channels of $T$. An important potential discovery channel which has not been explored is single production of $T$ followed by $T\rightarrow gt$ or $T\rightarrow \gamma t$.
 
 The production of single top quarks is now well enough understood that it has become possible to search for resonances in single top production, and to take advantage of the dominant all-hadronic decay mode. ATLAS studied the $t\gamma$ final state~\cite{ATLAS:2023qdu}, but did not examine the invariant mass of the system.

In this paper we study single production of a heavy vector-like quark $T$ which for high $T$ masses features higher cross sections than pair production, and study the all-hadronic fully reconstructable final states as well as the leptonic final states for the channels $T\rightarrow gt$ or $T\rightarrow \gamma t$.  
    
    The paper is organized as follows. The first section describes a benchmark model used to characterize the kinematics of resonant $tg$ or $t\gamma$ production via a heavy vector-like quark. The second section presents a study of the expected LHC sensitivity to these modes. The third section presents the results in terms of the theoretical parameters and discusses the wider context of possible models which could generate this signature.
\section{Models}
\label{sec:model}

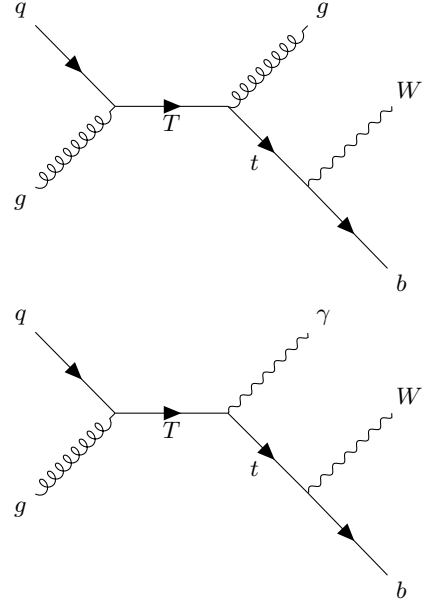
\begin{figure}[h!]
    \centering
     \begin{tikzpicture}
    \begin{feynman}
        \vertex (a) ;
        \vertex [above left=of a](i1){\(q\)};
        \vertex [below left=of a](i2){\(g\)};
        \vertex [right=of a] (b);
        \vertex [above right=of b] (f1) {\(g\)};
        \vertex [below right=of b] (c);
        \vertex [above right=of c] (f2) {\(W\)};
        \vertex [below right=of c] (f3) {\(b\)};
        \diagram* {
         (i1) -- [fermion] (a),
         (i2) -- [gluon] (a),
        (a) -- [fermion, edge label'=\(T\)] (b) -- [gluon] (f1),
        (b) -- [fermion, edge label'=\(t\)] (c),
        (c) -- [boson] (f2),
        (c) -- [fermion] (f3),
    };
    \end{feynman}
    \end{tikzpicture}
      \begin{tikzpicture}
    \begin{feynman}
        \vertex (a) ;
        \vertex [above left=of a](i1){\(q\)};
        \vertex [below left=of a](i2){\(g\)};
        \vertex [right=of a] (b);
        \vertex [above right=of b] (f1) {\(\gamma\)};
        \vertex [below right=of b] (c);
        \vertex [above right=of c] (f2) {\(W\)};
        \vertex [below right=of c] (f3) {\(b\)};
        \diagram* {
         (i1) -- [fermion] (a),
         (i2) -- [gluon] (a),
        (a) -- [fermion, edge label'=\(T\)] (b) -- [photon] (f1),
        (b) -- [fermion, edge label'=\(t\)] (c),
        (c) -- [boson] (f2),
        (c) -- [fermion] (f3),
    };
    \end{feynman}
    \end{tikzpicture}
    \caption{Feynman diagrams describing the single production of a heavy vector-like fermionic top partner $T$ which decays to a top quark and a gluon (top) or a top quark and a photon (bottom).}
    \label{fig:decay-diagram}
\end{figure}

Vector-like top partners can alleviate the hierarchy problem and are less constrained than a 4th generation of quarks with chiral couplings \cite{Aguilar_Saavedra_2013,Dobrescu:1997nm}. These particles can be found in more complete models such as Little Higgs models \cite{Schmaltz:2005ky}, supersymmetric theories, composite Higgs models \cite{Witzel:2019jbe} or in scenarios where the right-handed top quark is composite and $T$ is an excited state\cite{Lillie:2007hd,Delgado:2005fq}. However, the compatibility of these motivations and the model we study are dependent on the details of the underlying theory. In order to maximize model-independence, we search for $T$ through its direct couplings with SM fields at the LHC in such a way as to allow the results to be interpreted as broadly as possible. 
 
We consider a model which includes a new vector-like top partner $T$. Since $T$ and the right-handed top quark have the same quantum numbers, mixing can occur which leads to decays to electroweak bosons and top quarks.  In light of these constraints, we take an effective field theory approach and consider interactions that may be realized even in the event of zero mixing. We consider the dimension-5 operator
\be
{\cal{L}}_{EFT} \supset \frac{g^i_{\gamma}}{\Lambda} \bar{T}\sigma^{\mu\nu}u_{R,i}~ F_{\mu\nu} + \frac{g^i_{g}}{\Lambda} \bar{T}\sigma^{\mu\nu}u_{R,i}~ G_{\mu\nu}.
\label{EFTLagrangian}
\ee
Here, $u_{R,i}$ is a right-handed up-type quark of flavor $i \in \{1,2,3\}$, $F_{\mu\nu}$ ($G_{\mu\nu}=T^A G^A_{\mu\nu}$) is the photon (gluon) field strength tensor and $g^i_{\gamma}$ ($g^i_g$) is the flavor-dependent coupling, and the operator is suppressed by a scale $\Lambda$, which we set to 3 TeV in these studies.  While the interaction described above leads to $T$ pair production \cite{Alhazmi:2018whk} in $pp$ collisions, we consider the less explored option of mono-$T$ production.

These interaction terms allow for production of $T$ via quark-gluon fusion, and for decays such as $T\rightarrow qg$ and $\rightarrow q\gamma$. If the decay is to light-flavor quarks, dijet searches at the LHC place important constraints~\cite{CMS:2018wpl}, with an upper limit on the production cross section with $\sigma < 10$ fb for $m_T \sim 1$ TeV. 

\begin{figure}
    \centering
    \includegraphics[width=0.45\textwidth]{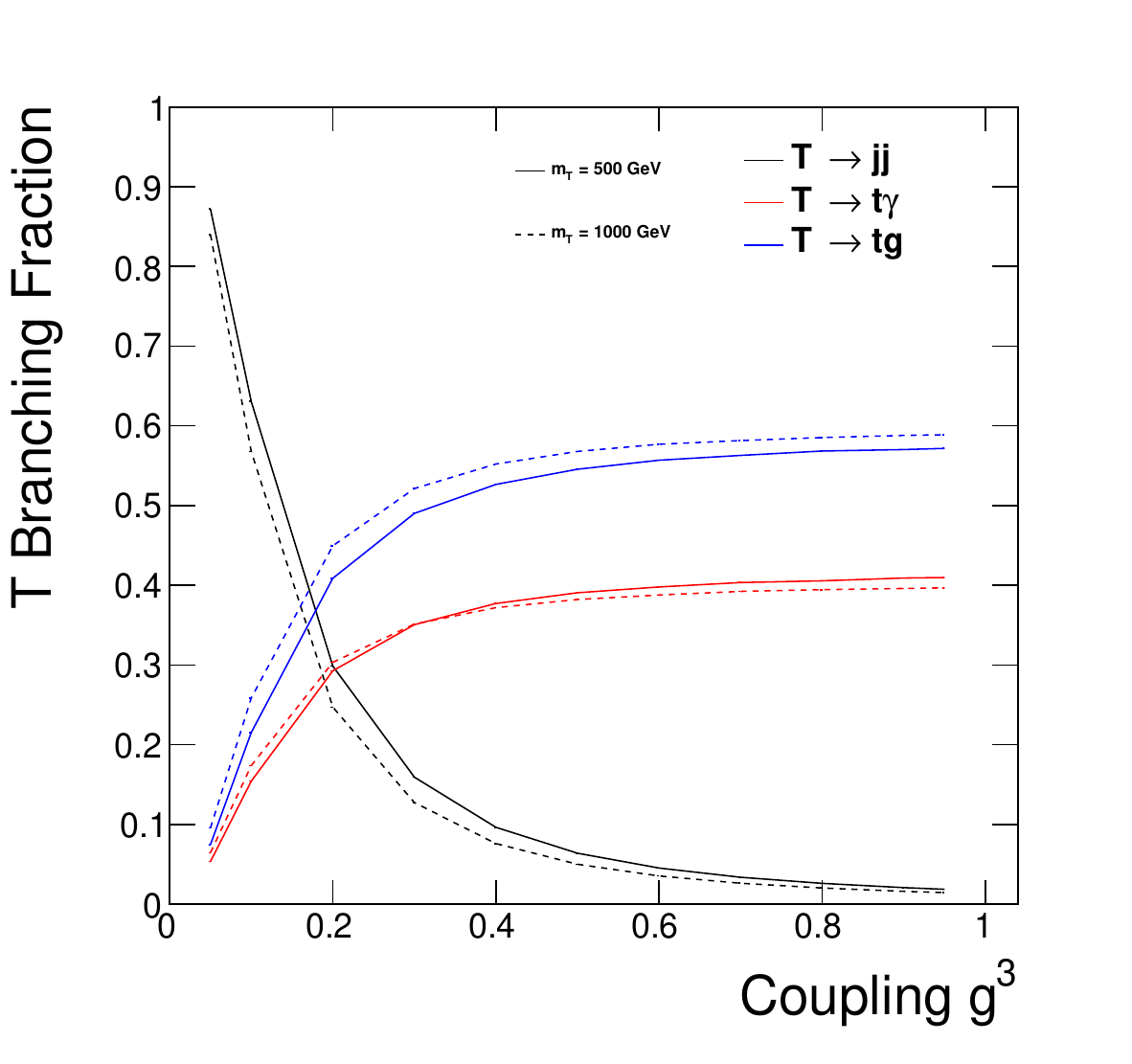}
    \caption{ Branching fraction of the heavy $T$ to light-flavor jets ($jj$), or to a top quark and photon ($t\gamma$) or gluon ($tg$), as a function of the coupling parameters $g_\gamma^3$ and $g_g^3$, varied together with $g^{1,2} = 0.1$.}
    \label{fig:bf}
\end{figure}

However, if the couplings are flavor-dependent, the production primarily depends on the light-quark coupling but the decay can be influenced by the heavy-flavor couplings, $g_\gamma^3 ,g_g^3$. Increased coupling to heavy flavor can result in decays predominantly to $tg$ and $t\gamma$; see Fig~\ref{fig:bf}. Though we do not study it here, it can also allow for $gg\rightarrow Tt$ production. For our $tg$ and $t\gamma$ analysis, we fix the coupling to light quarks $g_{\gamma,g}^{1,2} = 0.1$ and use decays to top quarks to constrain $g_{\gamma,g}^3$ (or equivalently the production cross section for $tg,t\gamma$ from $T)$. Of course, for $g_{\gamma,g}^3 \ll g_{\gamma,g}^{1,2}$, a dijet search will be the more sensitive channel. 

\section{Experimental Sensitivity}\label{sec:sens}

The models described above include interactions which can generate a final state with a single $T$ quark, as shown in Fig.~\ref{fig:decay-diagram}. We estimate the sensitivity of the LHC dataset to these hypothetical signals using samples of simulated $pp$ collisions at $\sqrt{s}=13$ TeV with an integrated luminosity of 300 fb$^{-1}$ over a mass range $m_T \in [150-1000]$ GeV. 


Simulated signal and background samples are used to reconstruct hypothetical resonances, estimate efficiencies and expected signal and background yields for $m_T\in [200 - 1000]~{\rm GeV} $. For the parameter space that we explore $\Gamma / M < 1\%$, such that the resonant production we consider in Fig.~\ref{fig:decay-diagram} is dominant and we do not need to consider additional non-resonant contributions to $T$ production \cite{Roy:2022wjj,Deandrea:2021vje}.
Collisions and  decays simulated with {\sc Madgraph5} v3.4.1 ~\cite{madgraph} for the matrix-element calculation are subsequently showered and hadronized with {\sc Pythia} v8.306~\cite{pythia}. The detector response is simulated with {\sc Delphes} v3.5.0~\cite{delphes} using the standard CMS card and {\sc root} version 6.26\/06 \cite{ROOT}, which includes a parameterized description of $b$-tagging.

Selected photons and leptons are required to have transverse momentum $p_\textrm{T}\geq10$ GeV and absolute pseudo-rapidity $0\leq|\eta|\leq2.5$. Selected jets are clustered using the anti-$k_{\textrm{T}}$ algorithm~\cite{Cacciari:2008gp} with radius parameter $R = 0.5$ using \textsc{FastJet 3.1.2}~\cite{Cacciari:2011ma} and are required to have $p_\textrm{T}\geq20$ GeV and $0\leq|\eta|\leq2.5$.   In addition, events are required to satisfy a lepton, photon or jet trigger requirement, as appropriate for each channel.

Below, we describe the selection and reconstruction in the $t\gamma$ and $tg$ modes.

\begin{table}
    \centering
        \caption{Expected yields in 300 fb$^{-1}$ due to background processes in each of the four final states.  Backgrounds may also include production of either a photon and one or two additional jets; see text for details. Also shown is the expected yield for signal with $m_T=500$ GeV with assumed coupling of $g^3_g=0.1$ ($g^3_{\gamma} = 0.1 ~{\rm cos}(\theta_W)$) in the gluon (photon) mode, with $\Lambda=3$ TeV.}
            \label{tab:yield}
    \begin{tabular}{cccccc}
    \hline\hline
    & \multicolumn{2}{c}{$T\rightarrow t\gamma$} &&\multicolumn{2}{c}{$T\rightarrow tg$}\\ \cline{2-3} \cline{5-6}
    & $bjj\gamma$ & $b\ell\nu\gamma$ && $bjjj$ & $b\ell\nu j$\\
    \hline
         $W$ &  19.7$\times 10^3$ & 22.0$\times 10^3$  && 6.0$\times 10^6$  & 7.4$\times 10^6$  \\
         $Z$& 9.6$\times 10^3$  & 500 && 2.0$\times 10^6$  & 236$\times 10^3$  \\
         $t\bar{t}$ &1.9$\times 10^3$  &  803 && 895$\times 10^3$  & 1.3$\times 10^6$  \\
         $tb$ &37.3  &39.8 & &159$\times 10^3$  & 753$\times 10^3$  \\
         \hline
         All Bg. &31.2$\times 10^3$  &23.3$\times 10^3$ & &9.0$\times 10^6$ & 9.7$\times 10^6$  \\
         Signal  & 1.5$\times 10^3$  & 190  && 970 & 350 \\
             \hline\hline

    \end{tabular}
\end{table}

\subsection{$t\gamma$ mode}

In this section, we consider the $T\rightarrow t\gamma$ mode, which produces a $bjj\gamma$ or $b\ell\nu\gamma$ final state, depending on the $W$ boson decay. The dominant backgrounds are production of a heavy boson associated with a photon and jet without an intermediate top quark, $p p \rightarrow Wj\gamma $ and $p p \rightarrow Zj\gamma$.  We also model contributions from top pair production ($t\bar{t}\gamma$) as well as $tb\gamma$; the additional $b$ quark in $p p \rightarrow t\gamma b$ is necessary in the SM due to a lack of flavor-changing vertices such as $Ttg$ or $Tg\gamma$.  Radiation of additional gluons is modeled by {\sc pythia}.

In the $bjj\gamma$ final state, events are required to have exactly three jets, to suppress the large top quark pair background, with exactly one $b-$tagged jet, one photon and zero leptons. To satisfy typical trigger requirements~\cite{ATL-DAQ-PUB-2019-001}, the photon must exceed 145 GeV. The two non-tagged jets correspond to the $W$ boson, and the three jets together correspond to the top quark. The mass of the $T$ quark, $m_T$, can be reconstructed from the three jets and the photon. Events with $m_T$ less than 200 are rejected.

In the $b\ell\nu\gamma$ final state, events are required to have exactly one photon, one lepton, one $b$-tagged jet, and missing transverse momentum of at least $25$ GeV. To satisfy typical trigger requirements~\cite{ATL-DAQ-PUB-2019-001}, the transverse momentum of the photon or lepton must exceed 145 GeV or 27 GeV respectively. The transverse momentum of the neutrino is assumed to be the missing transverse momentum. The longitudinal momentum $p_z^\nu$ is calculated under the assumption that the invariant mass of the $\ell\nu$ pair should be that of $m_W=80.3$~GeV. A $W$ boson candidate is reconstructed using the $\ell\nu$ pair and matched with the jet to build the top quark candidate. The $T$ quark candidates are reconstructed using photon and the top quark candidate. Events with $m_T$ less than 200 are rejected.

\begin{figure}
    \centering
        \includegraphics[width=0.4\textwidth]{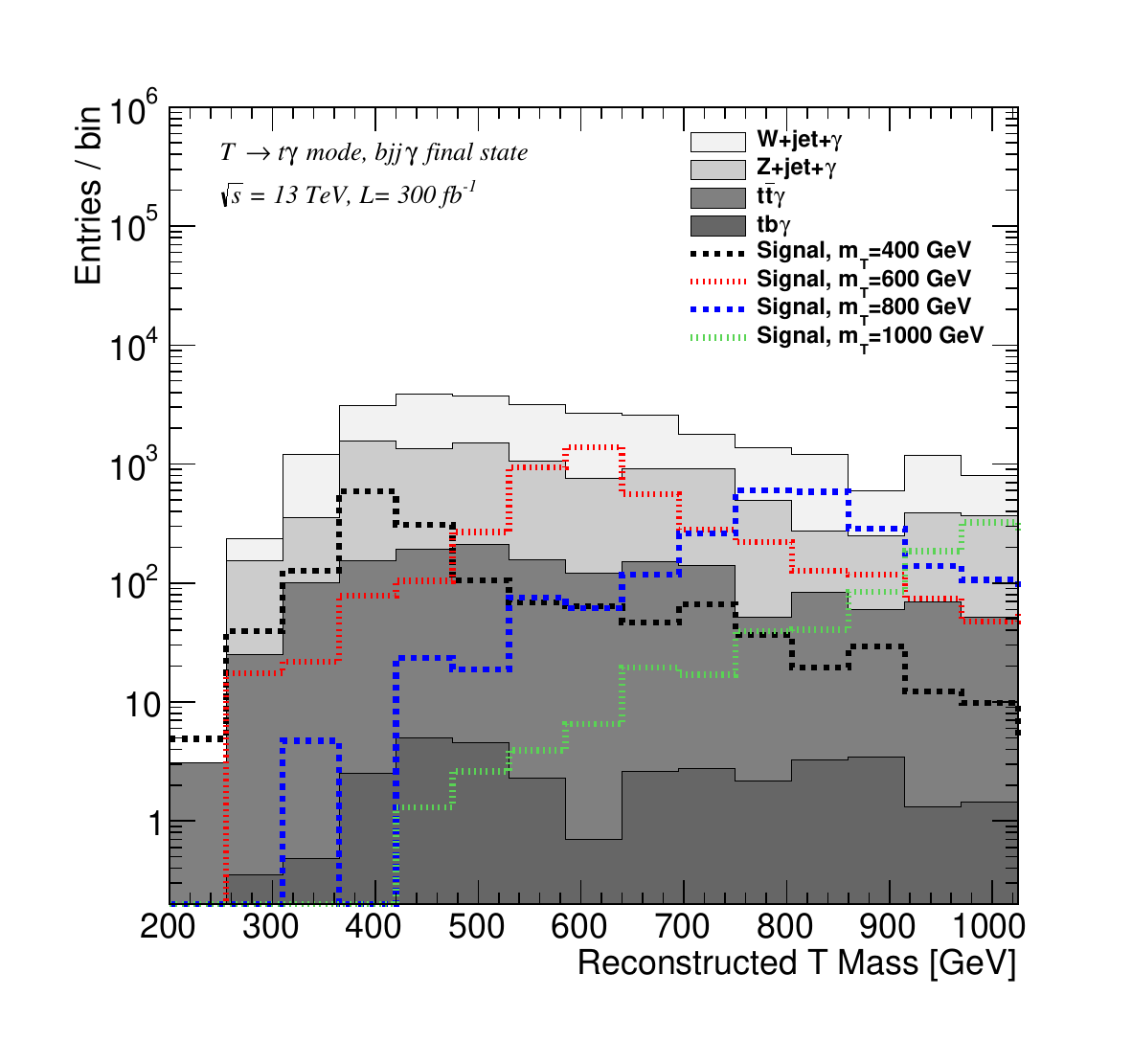}
        \includegraphics[width=0.4\textwidth]{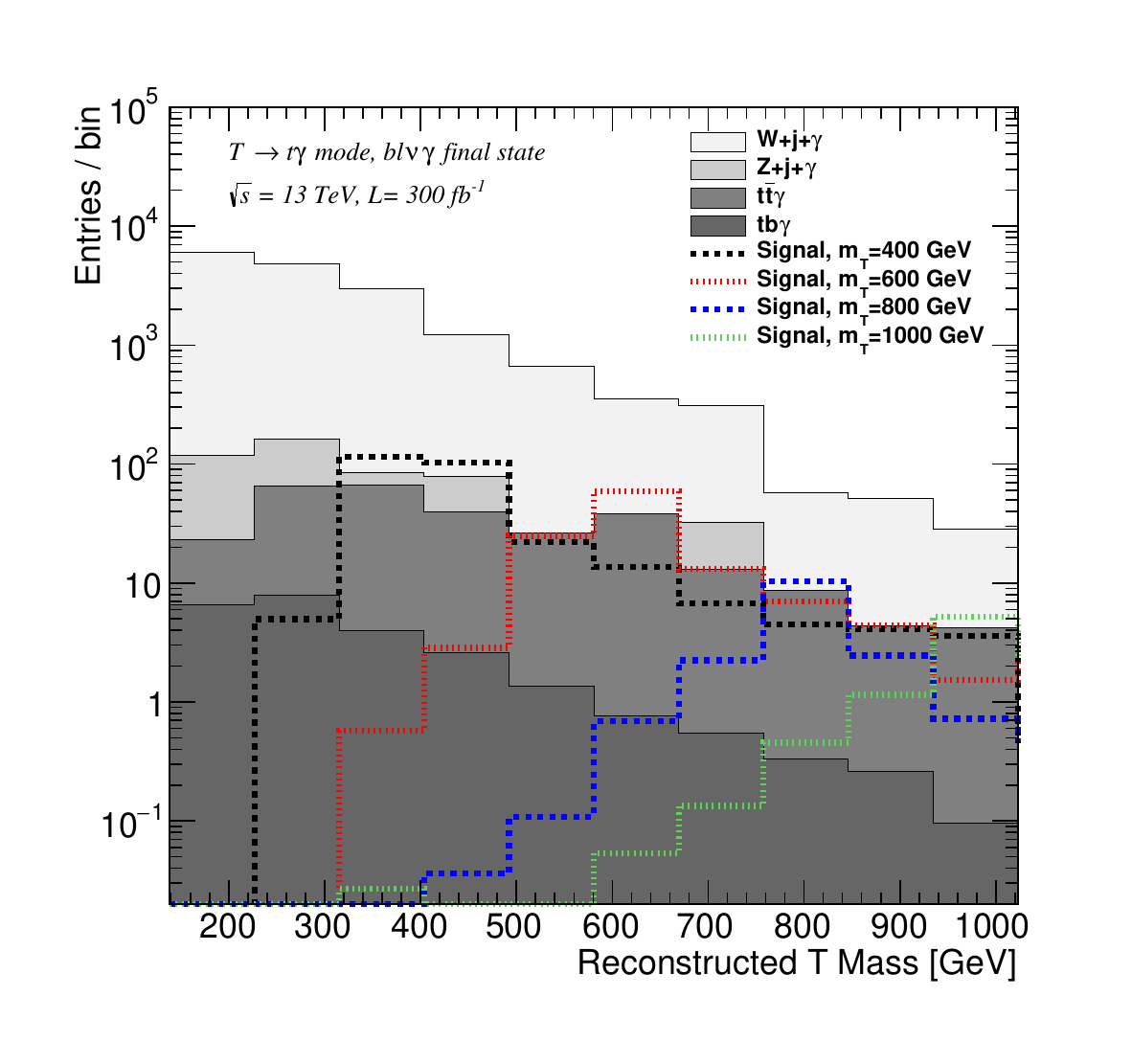}
    \caption{ Distributions of reconstructed $m_T$ in simulated signal and background samples for hadronic (top) and leptonic (center) decays in the $t\gamma$ mode,  normalized to an integrated luminosity of 300 fb$^{-1}$. The signal cross section is set to the expected limit at 95\% CL. }
    \label{fig:tgamma_plots}
\end{figure}

The distributions of reconstructed $m_T$ for signal and background samples are shown in Fig.~\ref{fig:tgamma_plots}.  Event yields for the various background processes as well as a hypothetical signal at $m_T = 500~{\rm GeV}$ are shown in Table \ref{tab:yield}.

\subsection{$tj$ final state}

In this section, we consider the $T\rightarrow tg$ mode, which produces a $bjjj$ or $b\ell\nu j$ final state, depending on the $W$ boson decay. The dominant backgrounds are production of a heavy boson associated with two jets without an intermediate top quark, $p p \rightarrow Wjj $ and $p p \rightarrow Zjj$.  We also model contributions from top pair production ($t\bar{t}$) as well as $tbj$; the additional $b$ quark in $p p \rightarrow tbj$ is necessary in the SM due to a lack of flavor-changing vertices such as $Ttg$ or $Tg\gamma$. Contributions from QCD production of $bjjj$ were found to be negligible above $m_T$ of 200 GeV. Other processes may also contribute to the background; rather than exhaustively catalog them, we take a 50\% systematic uncertainty in our background estimate. Radiation of additional gluons is modeled by {\sc pythia}. 

In the $bjjj$ final state, events are required to have exactly four jets, to suppress the large top quark pair background, with exactly one $b-$tagged jet, zero photons and zero leptons. To satisfy typical trigger requirements~\cite{ATL-DAQ-PUB-2019-001}, three jets must exceed 85 GeV or one jet must exceed 435 GeV.  The $b$-tagged jet is chosen as the candidate for the bottom quark and paired up with two non-$b$-tagged jets that give us the smallest $|m_W-m_{jj}|^2+ |m_t-m_{jjj}|^2$ value, where $m_{jj}$ is the invariant mass of the two non-$b$-tagged jets, and $m_{jjj}$ is the invariant mass of the three jets. The remaining non-$b$-tagged jet is then chosen to be the gluon candidate, and $T$ candidates are reconstructed using the top quark and gluon candidates. Events with a reconstructed $m_T$ greater than 200 GeV are selected.

In the $b\ell\nu j$ final state, events are required to have exactly  one lepton, one $b$-tagged jet, one non-tagged jet, and missing transverse momentum of at least $25$ GeV. To satisfy typical trigger requirements~\cite{ATL-DAQ-PUB-2019-001}, the lepton must exceed 27 GeV. Neutrinos are reconstructed with the same strategy employed in the $b\ell\nu\gamma$ final state. The top quark is reconstructed from the $\ell\nu$ pair and bottom quark. The remaining jet is selected to be the gluon candidate, and the $T$ particle is reconstructed using the gluon and top quark candidates.  Events with a reconstructed $m_T$ greater than 200 GeV are selected. 

\begin{figure}
    \centering
    \includegraphics[width=0.4\textwidth]{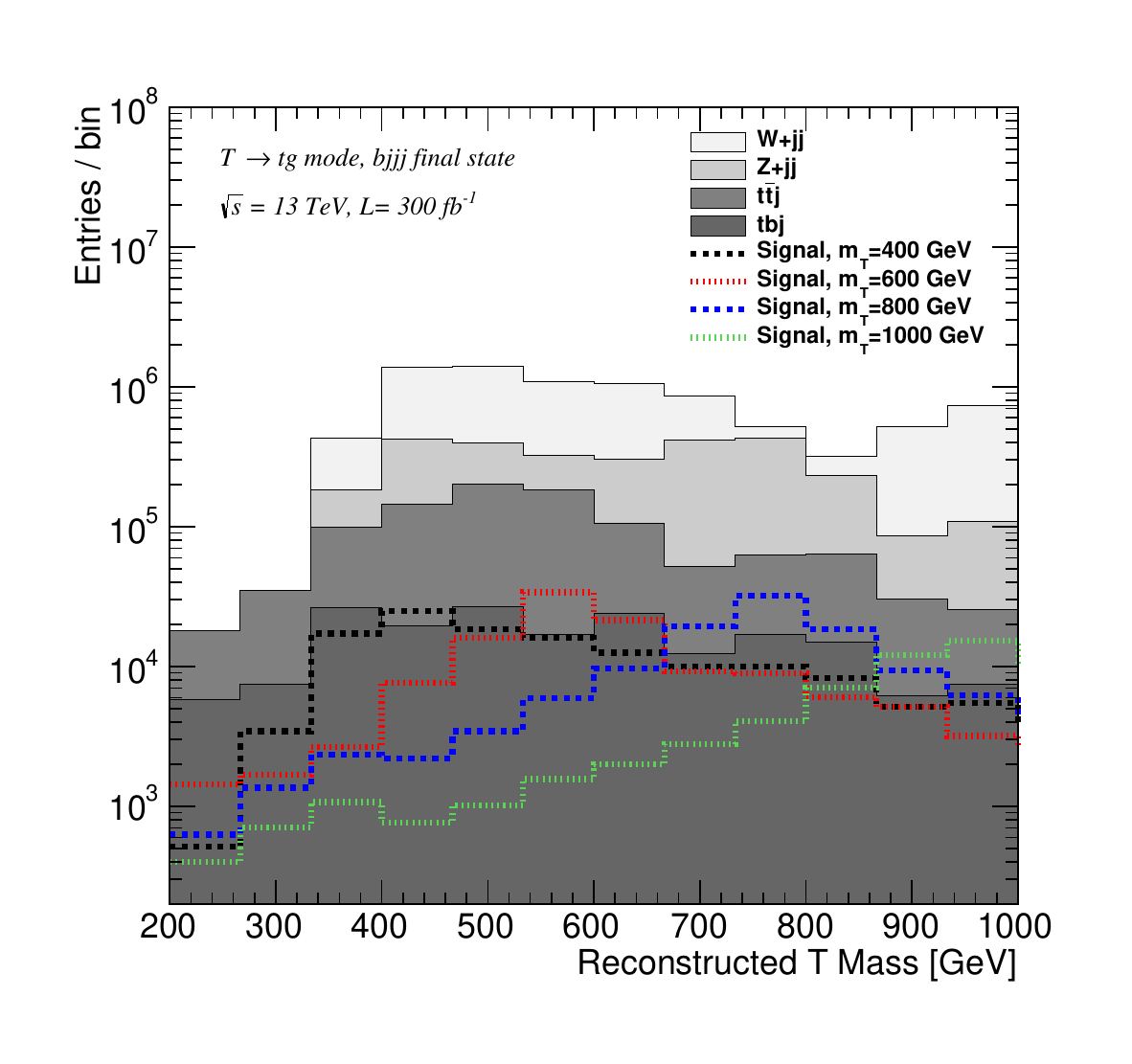}
    \includegraphics[width=0.4\textwidth]{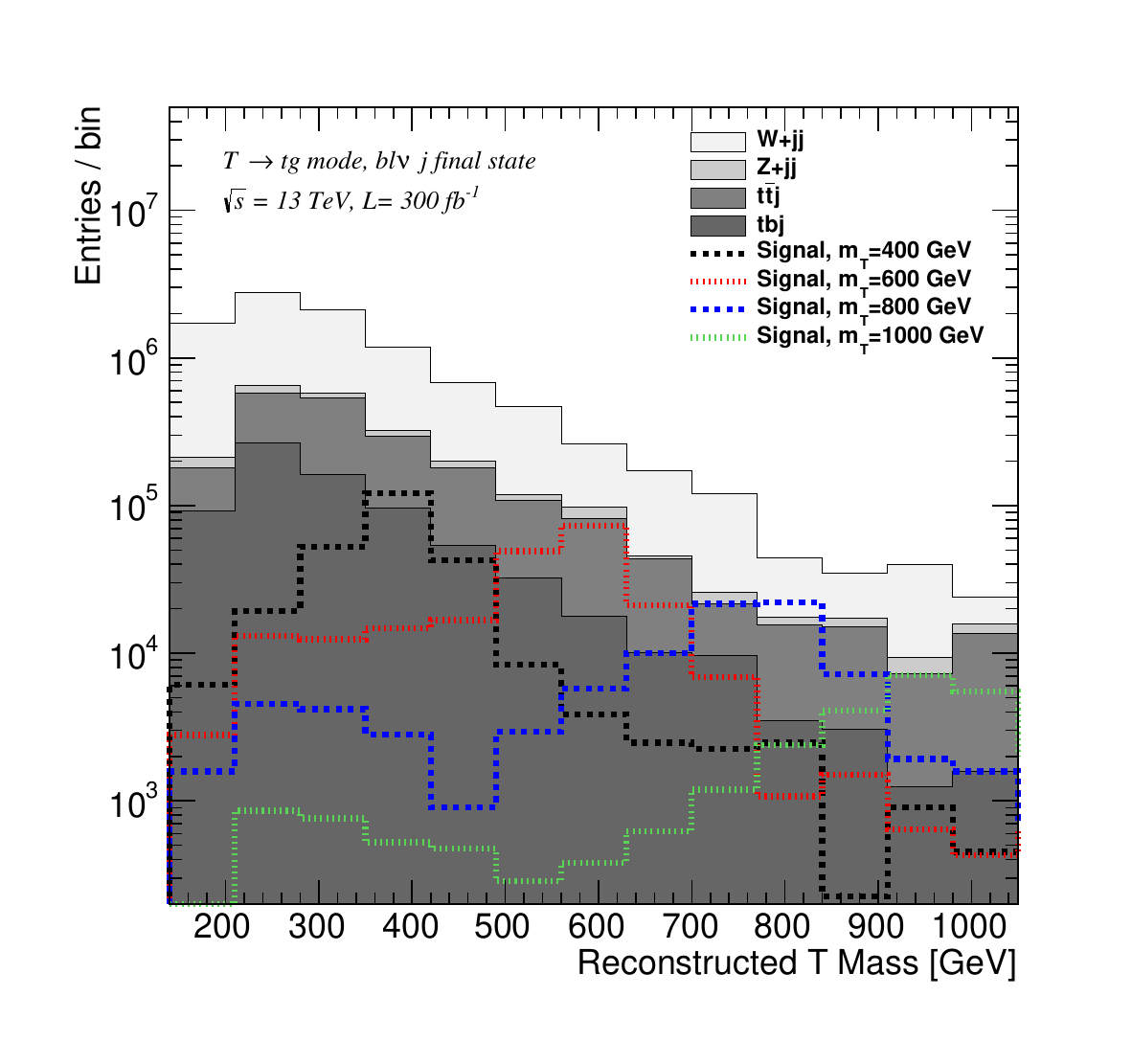}
    \caption{Distributions of reconstructed $m_T$ in simulated signal and background samples for hadronic (top) and leptonic (center) decays in the $tg$ mode,  normalized to an integrated luminosity of 300 fb$^{-1}$ for $\Lambda = 3$ TeV. The signal cross section is set to the expected limit at 95\% CL.}
    \label{fig:tjet_plots}
\end{figure}

The distributions of reconstructed $m_T$ for signal and background samples are shown in Fig.~\ref{fig:tjet_plots}. Event yields for the various background processes as well as a hypothetical signal at $m_T = 500~{\rm GeV}$ are shown in Table \ref{tab:yield}.
\subsection{Statistical Analysis}
\begin{figure}
    \centering
    \includegraphics[width = 0.5\textwidth]{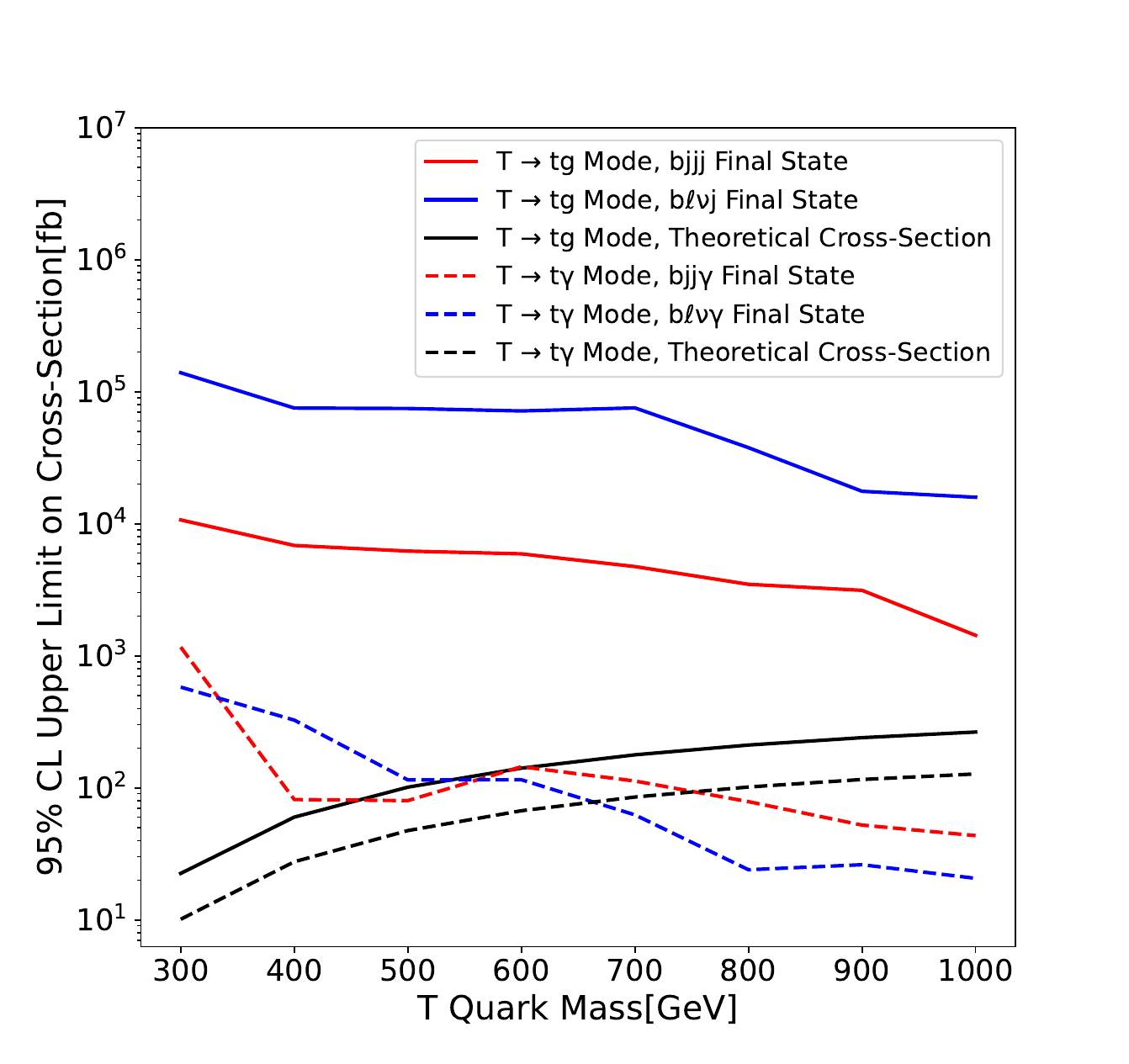}
    \caption{Top: summary of expected upper limits at 95\% CL on the production cross-sections as functions of the $T$ mass in integrated luminosity of 300 fb$^{-1}$ for the two final states and two modes. Also shown are expected theoretical cross sections at leading order with coupling values of $g^{1,2,3}_{\gamma,g}$ = 0.1 and $\Lambda = 3$ TeV. }
    \label{fig:xs_limit_summary}
\end{figure}

Limits are calculated at 95\% CL using a profile likelihood ratio~\cite{Cowan:2010js}  with the CLs technique~\cite{Junk:1999kv,Read:2002hq}. We use the pyhf~\cite{pyhf, pyhf_joss} package with a binned distribution in the reconstructed mass of the hypothetical $T$ boson, where bins without simulated background events have been merged into adjacent bins. The background is assumed to have a 50\% relative systematic uncertainty.  Limits as a function of the $T$ mass are shown in Figure~\ref{fig:xs_limit_summary}. 


\section{Discussion}
\label{sec:disc}

In the $t\gamma$ production mode, the leptonic decay  has a smaller efficiency due to the $W$ branching ratio, but also lower backgrounds.  The $T$ mass resolution is similar to that in the hadronic mode, as the lepton is well measured but the neutrino is unmeasured.  
In the $tg$ production mode, the hadronic decay benefits from increasing efficiency as the mass $m_T$ grows and overcomes the hadronic trigger requirements, which also suppress the large top quark pair background. As a result, the hadronic mode is nearly 10 times as sensitive as the leptonic mode, though both are order(s) of magnitude weaker in upper limits on the cross section than the $t\gamma$ mode, where the backgrounds are much smaller. Future studies may explore the potential advantages of wide-cone jets to better capture the boosted scenario for large $m_T$.

These limits depend only on the level of background and the kinematics of the $T\rightarrow tg$ and $T\rightarrow t\gamma$ production and decay kinematics, and are independent of assumptions about the total or relative production rates. In top-composite models, where the $T$ is an excited state of the top quark, the production rates can be significantly higher\cite{Lillie:2007hd,Delgado:2005fq}.  In this case there is expected to be an excess of four top production at the LHC~\cite{Lillie:2007hd,ATLAS:2015ktd,Pomarol:2008bh} as compared to the SM.

In addition to direct searches at the LHC, vector-like quarks in the ${\cal O}(100)$ GeV mass range can be searched for at the Tevatron~\cite{Okada:2012gy}. In the event of non-zero mixing between $T$ and $u_R$, electroweak precision observables can be used to constrain mixing parameters \cite{Chen_2014,Chen_2017,Dawson_2012,Choudhury:2001hs}. Furthermore,  flavor changing neutral current processes have been used to constrain vector-like fermions \cite{Ishiwata_2015,Aaboud_2018,DELAGUILA1985243,Kang_2019}.

\section{Conclusions}
\label{sec:conc}
In this paper we derive projected constraints for the production of a vector-like top partner with mass $m_T\sim{\cal O}(100)~{\rm GeV}$ at the LHC for an integrated luminosity of 300 fb$^{-1}$. To complement existing searches which look for pair production, we focus on the less-explored single $T$ production and search for $T$ decay into $tg$ and $t\gamma$ final states.

Searches for $T\rightarrow tg$ and $T\rightarrow t\gamma$ are an under-explored area of new physics in the top sector, especially in the hadronic decay mode.  Single production of such a heavy object can often extend experimental sensitivity to higher masses, opening a potential discovery window.

In the $tg$ decay channel, we find that the $bjjj$ mode is the most sensitive, probing production cross sections of $~10^4$ fb. The $t\gamma$ decay channel has similar sensitivities for each of the $bjj\gamma,b\ell\nu\gamma$ final states, with greater statistical power overall; it constrains cross sections of $~100$ fb for $m_T = 300 $ GeV, and $~20$ fb for $m_T = 1$ TeV, comparable to that achieved by dijet searches. These modes provide additional statistically independent channels to complement dijet searches.

\section{Acknowledgments} DW and M.Fenton funded by the DOE Office of Science. The work of M.Fieg was supported by NSF Grant PHY-2210283 and was also supported by NSF Graduate Research Fellowship Award No. DGE-1839285.  The authors thank Tim Tait, Avik Roy Jeong Han Kim and KC Kong for useful comments.

\clearpage

\bibliography{stops}

\end{document}